\documentclass[12pt]{article}
\usepackage{lingmacros}
\usepackage{tree-dvips}
\usepackage{graphicx}
\usepackage{subcaption}
\usepackage{amsmath}
\usepackage{amssymb}
\usepackage{latexsym}
\usepackage{color}
\usepackage{fullpage}
\usepackage{authblk}

\begin{document}

\title{{Thermodynamics and Van der Waals Phase Transition of Charged Black Holes in Flat Space via R\'enyi Statistics}}

\author[1,2,3]{Chatchai Promsiri \footnote{Email: chatchaipromsiri@gmail.com}} 
\affil[1]{\small Department of Physics, Faculty of Science, King Mongkut’s University of Technology Thonburi, Bangkok 10140, Thailand}
\affil[2]{\small Theoretical and Computational Physics (TCP), 
	Faculty of Science, King Mongkut's University of Technology Thonburi, 
	Bangkok 10140, Thailand}
\affil[3]{\small Theoretical and Computational Science Center(TaCS),
	Faculty of Science, King Mongkut’s University of Technology Thonburi, Bangkok 10140, Thailand}

\author[2,3,4]{Ekapong Hirunsirisawat \footnote{Email: ekapong.hir@mail.kmutt.ac.th}}
\affil[4]{\small Office of Engineering Science Classroom (ESC), Learning Institute, King Mongkut’s University of Technology Thonburi, Bangkok 10140, Thailand}

\author[1,2,3]{Watchara Liewrian \footnote{Email: watchara.liewrian@mail.kmutt.ac.th}}

\maketitle{}

\begin{abstract}
The phase structure and critical phenomena of the 3+1 dimensional charged black holes in asymptotically flat spacetime are investigated in terms of thermodynamic properties within the R\'enyi statistics. With this approach as the non-extensive parameter above zero, we find that the charged black hole can be in thermodynamic equilibrium with surrounding thermal radiation, and have a Hawking-Page phase transition in the same way in the case of AdS charged black hole.  This gives more evidence supporting the proposal that there exists an equivalence between the black hole thermodynamics in asymptotically flat spacetime via R\'enyi statistics and that in asymptotically AdS spacetime via Gibbs-Boltzmann statistics, proposed by Czinner et al.  However, the present work also provides another aspect of supporting evidence through exploring the extended phase space within the R\'enyi statistics.  Working on a modified version of Smarr formula, the thermodynamic pressure $P$ and volume $v$ of a charged black hole are found to be related to the non-extensive parameter.  The resulting $P-v$ diagram indicates that the thermodynamics of charged black holes in asymptotically flat spacetime via R\'enyi statistics has the Van der Waals phase structure, equivalent to that in asymptotically AdS spacetime via Gibbs-Boltzmann statistics.
\end{abstract}

\section{Introduction and Motivations}

In general relativity, a black hole was initially expected to be a dark object as a result from the existence of event horizon inside which nothing can escape. However, according to black hole thermodynamics initiated by Bekenstein \cite{Bekenstein} and Hawking \cite{Hawking}, a black hole can have entropy and non-zero temperature. Later on, there have been further works, namely the laws of black hole mechanics, suggesting that the area and the surface gravity of black hole's event horizon correspond with entropy and temperature, respectively \cite{Bardeen}. Furthermore, the electrostatic potential and angular velocity at the event horizon can be treated as the chemical potential. Surprisingly, the mathematical analogy between the laws of black hole mechanics and of thermodynamics has led us to the notion that a black hole could behave as a thermal object.

Usually, a zero-charge black hole in an asymptotically flat spacetime background can only in the phase with negative heat capacity at an arbitrary temperature, such that it cannot be in thermal equilibrium with a heat bath of radiations.  However, a rich phase structure can be found in some spacetime backgrounds through thermodynamic stability analysis. For instance, in an asymptotically AdS space, there are possibly two branches of uncharged black holes, namely the small and large black holes. The small black hole phase has negative heat capacity implying that it is thermodynamically unstable. On the other hand, the large black hole phase has positive heat capacity, hence it is thermodynamically stable. However, these two black hole phases can exist only above a certain temperature, let call it $T_{min}$, here. Below $T_{min}$, the pure thermal radiation bath occupies the AdS space. Intriguingly, there is the Hawking-Page phase transition at the temperature $T_{HP}$ slightly above $T_{min}$. At the temperature in the range between $T_{min}$ and $T_{HP}$, the thermal radiation is mostly thermodynamically preferred over the small and large black hole phases. The large black hole phase turns out to be the most thermodynamically preferred at $T>T_{HP}$ \cite{HawkingPage}.

{There are some controversial issues for long time whether it is appropriate to apply the standard Gibbs-Boltzmann (GB) approach into the self-gravitating system.} Gravitation is a long-range attractive force, such that the average potential energy $\langle V\rangle$ between particles in a self-gravitating system is negative. Given that $\langle K\rangle$ is a positive value of average kinetic energy, we can use the virial theorem to show that the total average energy of the system $\langle U\rangle$ has negative value, namely, $\langle U\rangle = \langle K\rangle + \langle V\rangle = - \langle K\rangle$. It is well-known that the kinetic energy $\langle K\rangle$ is linear in the temperature $T$ up to some constants due to the equipartition theorem. As a consequence, we obtain the heat capacity as $C = d\langle U\rangle /dT < 0$ \cite{Lynden,Thirring}. Importantly, a negative heat capacity indicates that the self-gravitating system is thermodynamically unstable. This implies that the system cannot become thermodynamic equilibrium when it is in thermal contact with a heat bath. While this is a weird behavior of the self-gravitating system, it might be possible that it is an incorrect conclusion. A complete understanding of this situation has still be challenging issue. Generically, the standard Gibbs-Boltzmann (GB) statistical approach should be violated in the case of long-range interactions due to some clues such as the existence of a divergent partition function. This has been pointed out by Gibbs \cite{Gibbs} and later on by others \cite{Landsberg1,Landsberg2,Tsallis1} (see also \cite{Tsallis2} and references therein). In other words, applying the standard GB approach in the case of self-gravitating system may lead us to obtain an incomplete results.  

In conventional systems of ordinary matter in thermodynamics, the entropy of a whole system can be written as the sum of the entropy of subsystems. In this way, the entropy of the system typically scales with its volume. As a result, it is said to be an \textsl{extensive variable}. However, Bekenstein argued that a black hole system carries entropy proportional to the surface area of its event horizon rather than the volume. Such a behavior is so called {an} \textsl{area law}. {Therefore, the black hole's entropy is non-extensive.} Intriguingly, this black hole area law guides us to the holographic principle, which states that the information in a higher dimensional bulk spacetime can be endcoded into its boundary \cite{Hooft,Susskind1}. Later on, this principle becomes more established by the developments of the AdS/CFT correspondence \cite{Maldacena,Witten1,Witten2}.  {Moreover}, the area law can be found in the entanglement entropy of the reduced state of a subregion in strongly correlated quantum system \cite{Srednicki}.  This similarity raises the question about whether the quantum origin of black hole entropy may be somehow related to the entanglement entropy.  Currently, the non-extensive nature of entropy has received wide attention in several fields, and it might possibly improve our insights about microscopic nature of black hole.
 
As discussed above, the GB statistics might not be appropriate to use in black hole thermodynamics. In other words, the black hole entropy $S_{BH}$ should encode the black hole information with non-local and non-extensive nature. Consequently, we need non-Boltzmannian approach to deal with this. A new type of entropic function can be introduced by relaxing the Shannon-Khinchin axiomatic definition of the entropic function, \textit{i.e.} additivity, to the weaker non-additive composition rule. With composibility, the entropic function can be kept to be physically meaningful. According the derivation of Abe, the most general non-additive entropy composition rule is in the form \cite{Abe}
\begin{equation} 
H_{\lambda}\left(S_{12}\right)=H_{\lambda}\left(S_{1}\right)+H_{\lambda}\left(S_{2}\right)+\lambda H_{\lambda}\left(S_{1}\right) H_{\lambda}\left(S_{2}\right), \label{non additive}
\end{equation} 
where $H_{\lambda}$ is a differentiable function of $S$ and $\lambda \in \mathbb{R}$ is a constant parameter. One of the simplest version of non-extensive entropy, obeying Tsallis entropy, can be written in the form \cite{Tsallis1}
\begin{eqnarray}
S_T=\frac{1}{1-q}\left( \sum _{i=1}^W p_i^q-1 \right),
\end{eqnarray}
where $p_i$ are the probabilities of microstates of the system, W is the total number of microstates, and $q\in \mathbb{R}$ is the dimensionless parameter of non-extensivity. Clearly, the standard GB entropy is recovered when $q\rightarrow 1$. The composition rule of non-additive Tsallis entropy can be written as
\begin{eqnarray}
S_T^{12}=S_T^1+S_T^2+(1-q)S_T^1S_T^2, \label{Tsallis com}
\end{eqnarray}
which satisfies the Abe's non-additive entropy composition rule, as shown in \eqref{non additive}, when $H_\lambda(S)=S_T$ and $\lambda=1-q$. However, there is a long-standing problem about the thermal equilibrium {for non-extensive systems relating to the compatibility of the zeroth law of thermodynamics. Namely,} if two systems are each in thermal equilibrium then the total entropy has the maximum value $dS_{AB}=d(S_A+S_B)=0$, which implies the existence of an empirical temperature $\frac{1}{T}=\frac{\partial S_A}{\partial E_A}=\frac{\partial S_B}{\partial E_B}$. The problem of non-extensive entropy is that its composition rule is in the form of non-additive, hence it is not clearly whether  we can define an empirical temperature \cite{Nauenberg}. {Recently}, Bir\'o and V\'an proposed a way to solve this problem by transforming the non-additive entropy into another one that have an additive composition rule, which satisfies the zeroth law of thermodynamics \cite{Biro}. Their method is so called the \textsl{formal logarithmic approach}. They showed that, for homogeneous system,  the Tsallis entropy can be transformed into the well-defined entropy function as
\begin{eqnarray}
L(S_T)=\frac{1}{1-q}\left[ \ln (1+(1-q)S_T) \right] \equiv S_R.
\end{eqnarray}
Interestingly, this result is a well-known R\'enyi entropy $S_R$, which can be defined up to an arbitary real parameter $q$ as \cite{Renyi}
\begin{eqnarray}
S_R=\frac{1}{1-q}\ln \sum _{i=1}^W p_i^q.
\end{eqnarray}
For $q$ approaches to 1, it reduces to standard GB entropy. Notice that the composition rules of the R\'enyi entropy for independent subsystems are in additive {form
\begin{eqnarray}
S_R^{12}&=&S_R^1+S_R^2.
\end{eqnarray}
Now,} the entropy function is compatible with the zeroth law of thermodynamics and then we can uniquely defined an empirical temperature in thermal equilibrium between subsystems as
\begin{eqnarray}
\frac{1}{T_R} = \frac{\partial S_R(E)}{\partial E},
\end{eqnarray}
where $E$ is the energy of the system.

The non-Boltzmannian approach has been applied to investigate the black hole entropy problem in \cite{Tsallis3} and references therein. Recently, the thermodynamic stability of Schwartzchild and Kerr black holes in asymptotically flat spacetime were investigated via R\'enyi statistics in \cite{Czinner1,Czinner2}. {Interestingly, their works have shown that the black holes in asymptotically flat spacetime can be in stable equilibrium with the heat bath at a fixed temperature ensemble at the non-extensive parameter $\lambda = 1-q$ above zero. Moreover, the black holes in R\'enyi model can have a small/large black hole first order phase transition in the same way as the Hawking-Page phase transition of black holes in AdS space by using the conventional GB statistics.  Although the connection between these two pictures has been shown and claimed in previous works, this cannot be shown in a convincing way. It is interesting to raise a question about how the $\lambda$ parameter relates to the cosmological constant $\Lambda$ of AdS space. We will explore about this in the case of charged black hole.}  

{In conventional thermodynamics, the phase structure of the asymptotically AdS Reissner-Nordstr\"om black holes (RN-AdS) has been investigated in both grand canonical (fixed potential) and canonical (fixed charged) ensembles, see  
\cite{Chamblin1,Chamblin2,Chatchai}. In particular, RN-AdS black holes in a canonical ensemble have been found to thermodynamically behave in a similar way as the Van der Waals (VdW) liquid-gas system.}  Recently, more concrete comparison can be acheived by identifying thermodynamic pressure $P$, specific volume $v$ and temperature $T$ in VdW system with the cosmological constant $\Lambda$, the black-hole horizon radius $r_+$ and the Hawking temperature  $T_H$ of RN-AdS black hole, respectively. We can treat the RN-AdS black hole as a chemical system with  
\begin{eqnarray}
P  =  -\frac{\Lambda}{8\pi} = \frac{3}{8\pi l_{AdS}^2} , \ \ \ v  =  2l_p^2r_+ , \ \ \ T =  T_H. \label{p_extend}
\end{eqnarray}
The consideration in this way is so called the \textsl{extended phase space} approach \cite{Kastor,Mann1,Mann2}.

In the present paper, motivated by non-extensive nature of black hole entropy and VdW-like phase structure of charged black hole in the AdS background in the extened phase space approach, we explore the thermodynamics of 3+1 dimensional Reissner-Nordstrom black hole in asymptotically flat spacetime (RN-flat) through R\'enyi statistics and then investigate its phase structure in extended phase space. To acheive this, we need to {modify the Smarr relation in terms of R\'enyi entropy and its coresponding empirical temperature. With this approach, the thermodynamic pressure is found to relate with the non-extensive parameter.} This is reminiscent of the relation between thermodynamic pressure $P$ and cosmological constant $\Lambda$ in an extended phase space approach as shown in \eqref{p_extend}. Furthermore, the VdW-like phase transition has also been found in this setting. These give us more strong evidences about the connection between thermodynamics of {black holes in AdS space calculated by GB statistics and black holes in flat space calculated by R\'enyi statistics.} We discuss about this connection in this paper.  

The organization of the paper is as follows. In Section 2, we review the spherical charged black hole solution in asymptotically flat spacetime and {discuss} its standard thermodynamics for fixed potential and fixed charge ensemble. In Section 3, the thermodynamic properties of the RN-flat are investigate within the R\'enyi statistics.  The thermal phase structure is also studied in this section. Then, we suggests in Section 4, an analogous thermal behavior of the charged black hole in R\'enyi statistics to the Van der Waals liquid-gas system. We conclude the results and give suggestion of further studies in Section 5.

\section{Review of Charged Black Hole Thermodynamics}

In this section, we review the standard thermodynamics of the 3+1 dimensional RN-flat spacetime. Starting with a spherical symmetric Reissner-Nordstr\"om metric of the mass $M$ and the charged $Q$ in the form
\begin{eqnarray}
ds^2 = -f(r)dt^2 + \frac{dr^2}{f(r)} + r^2d\Omega^2_2,
\end{eqnarray}
where $d\Omega^2_2 = d\theta^2 + \sin^2\theta d\phi^2$ is the square of line element on 2-sphere and the function $f(r)$ is given by
\begin{eqnarray}
f(r)=1-\frac{2M}{r} + \frac{Q^2}{r^2}.
\end{eqnarray}
The black hole horizon can be determined from the condition $f(r)=0$, where its roots consist of  
\begin{eqnarray}
r_{\pm}=M \pm \sqrt{M^2-Q^2},
\end{eqnarray}
where $r_+$ and $r_-$ are the radius of the outer and inner horizons, respectively.   The black hole horizon is at $r_+$, from which the Hawking radiations are generated.  The black hole mass $M$ relates to event horizon radius and charge $Q$ as follows
\begin{eqnarray}
M = \frac{r_+}{2}\left( 1 + \frac{Q^2}{r_+^2} \right). \label{mass}
\end{eqnarray}
The charge $Q$ can generate the bulk gauge field in the form  
\begin{eqnarray}
A = A_tdt = -\left(\frac{Q}{r}-\Phi \right)dt. \label {bh4}
\end{eqnarray}
By setting $A_t=0$ at the horizon, we can relate the electric potential $\Phi$ with $M$ and $Q$ using $r_+ = M+\sqrt{M^2-Q^2}$. Thus, we have
\begin{equation}
\Phi = \frac{Q}{r_{+}} = \frac{Q}{M + \sqrt{M^2 - Q^2}}. \label {bh6}
\end{equation}
Obviously, the outer and inner horizons are degenerate when $M=Q$.  Namely, this extremal condition of black hole has only one horizon, $i.e.$ $r_e = r_+ =r_-$.  On the other hand, the charged balck hole have no horizon when $M<Q$.  To avoid this condition of naked singularity, the mass of the black hole must be not smaller than its charge, i.e., $M \geq Q$, therefore the electric potential is allowed to have the value in the range $0 \leq \Phi \leq 1$.

In black hole thermodynamics, physical quantities in a black hole system can be treated as thermodynamic variables. This results from the mathematical analogy between the laws of black hole mechanics and the laws of thermodynamics. The first law of black hole mechanics relates two stationary black holes with the change in mass  $\delta M$, the horizon surface area $\delta A$, the angular momentum $\delta J$  and electric charge  $\delta Q$ in the form \cite{Bardeen}
\begin{eqnarray}
\delta M = \frac{\kappa}{8\pi G}\delta A + \Omega \delta J + \Phi \delta Q,
\end{eqnarray}
where $\kappa$, $\Omega$ and $\Phi$ are surface gravity, angular velocity and electric potential at the event horizon, respectively. This can be seen as the first law of thermodynamics when one identifies the event horizon area $A$ and surface gravity $\kappa$ with entropy and temperature of the black hole, respectively. For non-rotating charged black hole, $Q$ is the number of particles in thermodynamic description, since the charge simply counts the number of particles and its conjugate $\Phi$ plays a role of chemical potential.

\subsection{Grand Canonical Ensembles}

One can consider thermodynamics of black hole in either grand canonical ensemble or canonical ensemble. When the black hole exchanges charge $Q$ with a surrounding heat bath, the chemical potential $\Phi$ can be held to be fixed. In this way, the system is being considered in the grand canonical ensemble.  Consequently, the charged black hole thermodynamic quantities can be written as
\begin{eqnarray}
E &=& M = \frac{r_+(1+\Phi^2)}{2}, \\
T_H &=& \frac{f'(r_+)}{4\pi} = \frac{1-\Phi^2}{4\pi r_+}, \\ \label {bh10} 
S_{BH} &=& \frac{A}{4} = \pi r_+^{2}, \\ \label {bh11}
C_{\Phi} &=& T_H\left( \frac{\partial{S_{BH}}}{\partial{T_H}} \right)_\Phi = -\frac{2\pi r_+^{2}(1+\Phi^2)}{1-\Phi^2}, \\ \label{bh13}
G &=& E - T_HS_{BH} - \Phi Q = \frac{r_+(1-\Phi^2)}{4},
\end{eqnarray}  
where $E$ is the internal energy, $T_H$ is the Hawking temperature, $S_{BH}$ is the Bekenstein-Hawking entropy, $C_\Phi$ is the heat capacity in a fixed electric potential $\Phi$ case and $G$ is the Gibbs free energy, respectively. Because the heat capacity $C_{\Phi}$ is negative at an arbitrary value of $r_+$ and $\Phi$, therefore  the black hole phase cannot be in a stable equilibrium with pure thermal radiation in the grand canonical ensemble.  This is shown as black dashed line in Fig.~\ref{fig:1} and \ref{fig:2}.

\subsection{Canonical Ensembles}

Working in the canonical ensemble, we consider a black hole of fixed charge $Q$.  In this case, we measure the energy $E$ of the system with respect to the ground state which is the extremal black hole. The mass $M_e$ of the extremal black hole can be obtained from the condition that the $r_+$ (outer) and $r_-$ (inner) horizon radius are degenerated. Hence the mass and charge of an extremal black hole are expressed as $M_e = Q = r_e$. Therefore the thermodynamic quantities in the canonical ensemble take the forms
\begin{eqnarray}
E &=& M - M_e = \frac{(r_+ - Q)^2}{2r_+}, \\
T_H &=& \frac{f'(r_+)}{4\pi} = \frac{r_+^2 - Q^2}{4\pi r_+^3}, \label{Thq} \\
S_{BH} &=& \frac{A}{4} = \pi r_+^{2}, \\
C_Q &=& T_H\left( \frac{\partial{S_{BH}}}{\partial{T_H}} \right)_Q=-\frac{2\pi r_+^{2}(r_+^{2}-Q^2)}{r_+^{2} - 3Q^{2}}, \label{Cq} \\
F &=& E - T_HS_{BH} = \frac{(r_+ - Q)(r_+ - 3Q)}{4r_+},
\end{eqnarray}
where all  thermodynamics variables are defined as in the grand canonical ensemble except that $C_Q$ here, is heat capacity at constant number of particles and $F$ is the Helmholtz free energy. From \eqref{Cq}, we have two branches of charged black holes in which each has either positive or negative heat capacity. The heat capacity is positive when 
\begin{eqnarray}
Q < r_+ < \sqrt{3}Q,
\end{eqnarray}
whereas it is negative when $r_+ > \sqrt{3}Q$. At the critical point $r_c=\sqrt{3}Q$, the heat capacity diverges as the Hawking temperature reaches its maximum value
\begin{eqnarray}
T_{max}= \frac{1}{6\sqrt{3}\pi Q}. \label{bh15}
\end{eqnarray}
In contrast with typical systems, the phase transition of RN-flat black holes depends on the ensemble in consideration. This might result from the long-range nature of gravitational and electromagnetic interactions of the system \cite{Bouchet}.

\section{Thermodynamics and Thermal Phase Transition via R\'enyi Statistics}

Now, we turn to an alternative approach to study thermodynamic properties of charged black holes. Here, we use the R\'enyi statistics and then calculate thermodynamic quantities. Then, we will investigate thermal phase structure of RN-flat black holes in both fixed potential and fixed charge ensembles.

As suggested in previous works \cite{Czinner1,Czinner2}, black holes can be treated in the way that it follows the non-additive Tsallis statistics, whose the composition rule is in the simplest form as shown in \eqref{Tsallis com}. Even though the Tsallis entropy function tends to be one of the proper choices representing black hole {entropy}, it turns out to have a difficulty to define the empirical temperature through the zeroth law of thermodynamics. To avoid this, the R\'enyi entropy in the form of the formal logarithm of the Tsallis entropy is proposed \cite{Biro}. Using this transformation rule, we obtain the R\'enyi entropy function of a black hole as

\begin{equation}
S_R=\frac{1}{\lambda}\ln(1+\lambda S_{BH}), \label{bh17}
\end{equation}
which has an additive property. The R\'enyi temperature can be expressed as
\begin{eqnarray}
T_R = \frac{1}{\partial{S_R/\partial{M}}} = T_H(1+\lambda S_{BH}). \label{Tr}
\end{eqnarray}
Based on these formula of the R\'enyi entropy and its corresponding temperature, we consider in this section the thermal phase transitions of the RN-flat black holes in the grand canonical and the canonical ensemble. 

\subsection{Grand Canonical Ensemble}

For the grand canonical ensemble with a fixed value of $\Phi$, the corresponding R\'enyi temperature as a function of the event horizon $r_+$ and the electrical potential $\Phi$ is given by
\begin{eqnarray}
T_{R}  =  \frac{(1-\Phi^2)(1+\lambda \pi r_+^{2})}{4\pi r_+}. \label{Tphi}
\end{eqnarray}
The heat capacity can be obtained in a usual way   
\begin{eqnarray}
C_{R}&=&T_{R}\left( \frac{\partial S_{R}}{\partial T_{R}} \right)_{\Phi} = -\frac{2\pi r_+^{2}(1+\Phi^2)}{(1-\Phi^2)(1 -  \lambda \pi r_+^{2})}. \label{bh21}
\end{eqnarray}
Obviously, the heat capacity is negative when $r_+ < r_{c}$ and positive when $r_+ > r_{c}$, where  $r_c=\sqrt{1/\lambda \pi}$.  Accordingly, there are two possible black hole configurations: one with negative value of heat capacity and another with positive value. We will refer to these as small and large black holes, respectively. Substituting $r_c$ into \eqref{mass}, we obtain the critical mass
\begin{eqnarray}
M_c = \frac{1+\Phi^2}{2\sqrt{\lambda \pi}}.
\end{eqnarray}
In R\'enyi statistics, a RN-flat black hole is in the small black hole branch when $M<M_c$ and in the large black hole branch when $M>M_c$.

\begin{figure}
 	\centering
 	 \begin{subfigure}[b]{0.43\textwidth}
 	  \includegraphics[width=\textwidth]{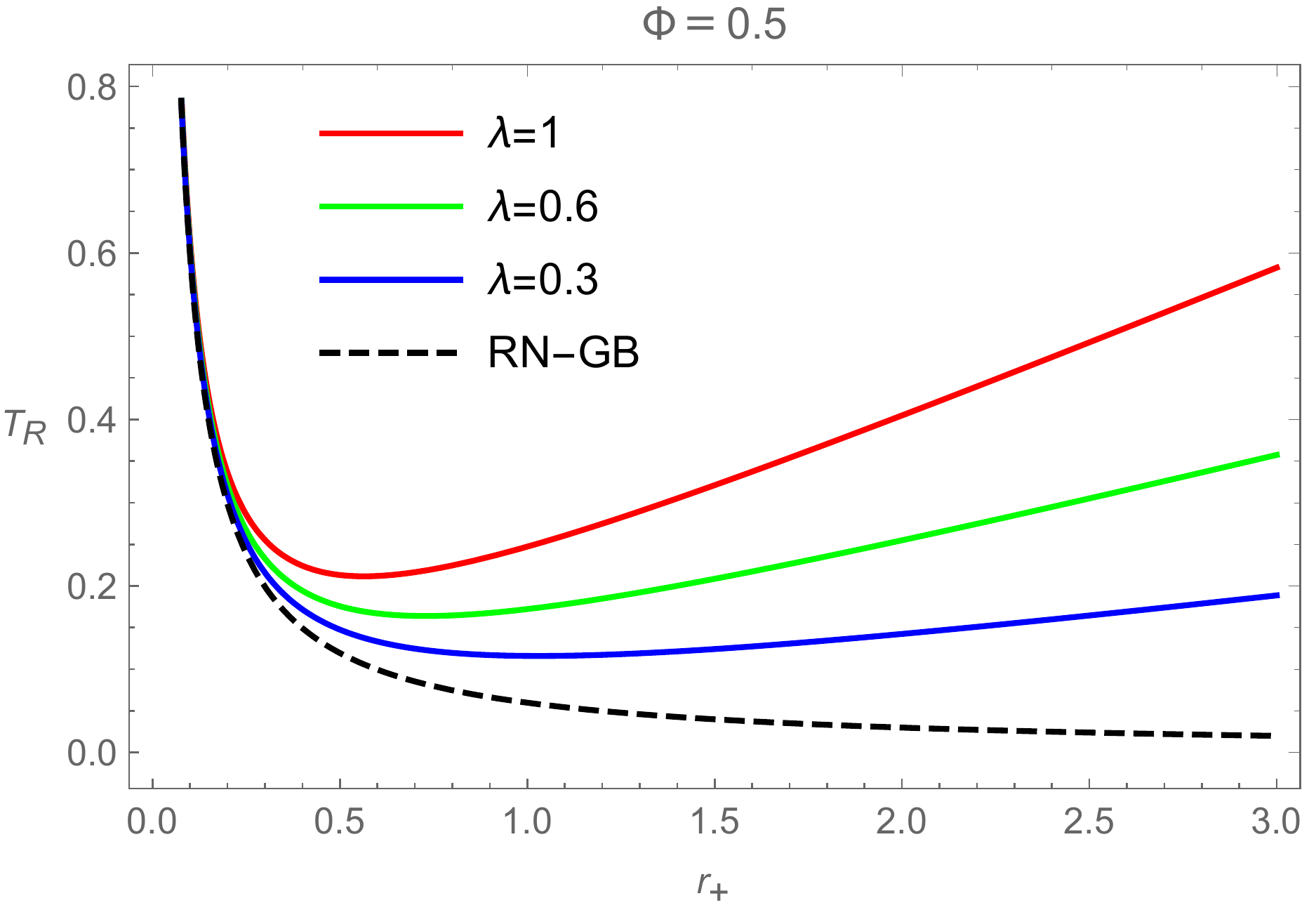}
 	   \end{subfigure}
 	   \begin{subfigure}[b]{0.44\textwidth}
 	    \includegraphics[width=\textwidth]{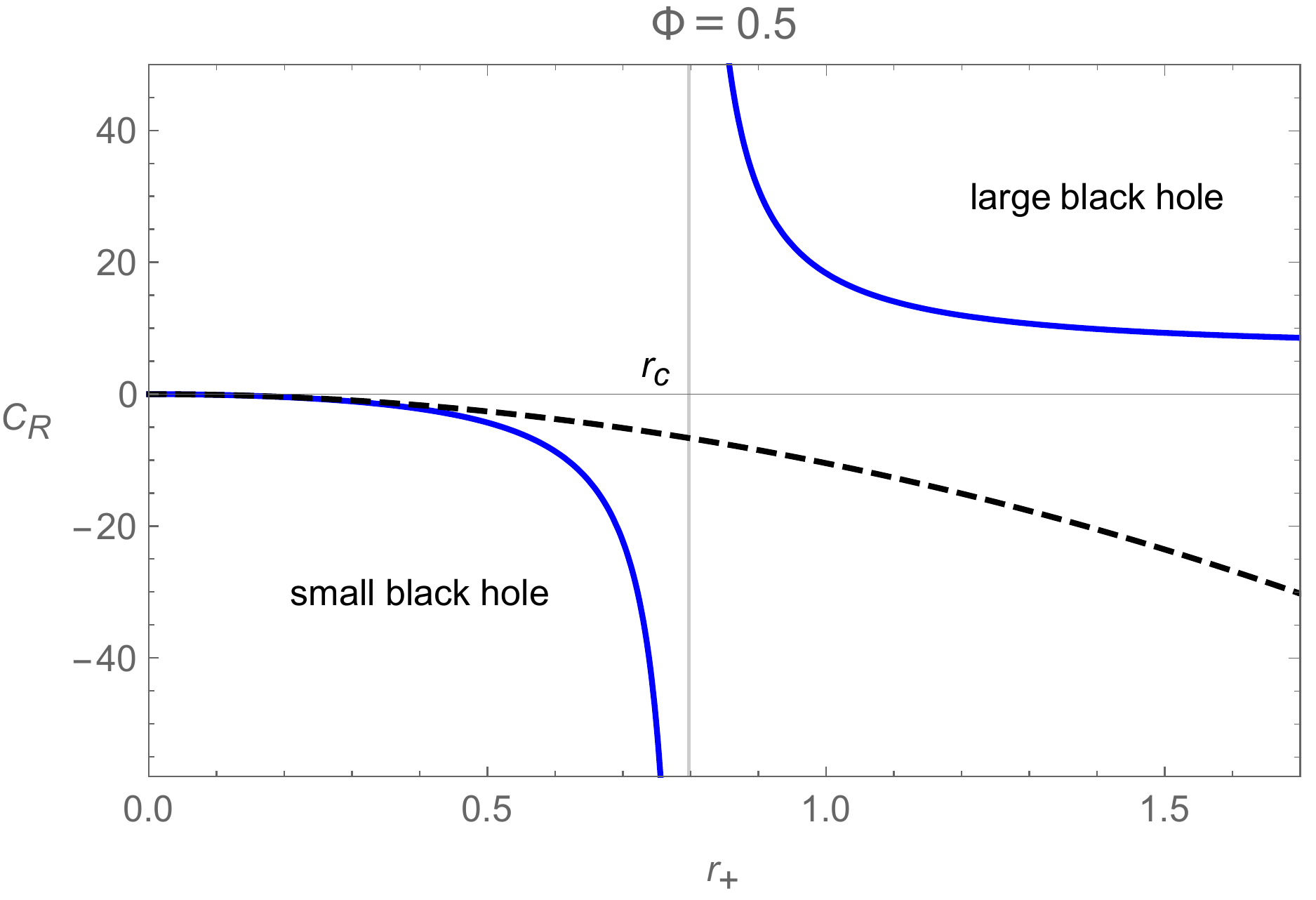}
 	     \end{subfigure}
 	      \caption{(left) The R\'enyi temperature of a charged black hole $T_R$ versus the event horizon radius $r_+$ for a fixed value $\Phi = 0.5$ is plotted with $\lambda =$ 0.3 (solid blue), 0.6 (solid green) and 1.0 (solid red), compared with the case of the GB statistics (dashed black). (right) The heat capacity $C_R$ of a charge black hole versus $r_+$ with $\Phi = 0.5$ is plotted in the case of $\lambda = 0.3$ (solid blue), its value is negative at $r_+ < r_c$, while it is positive at $r_+>r_c$. However, the heat capacity cannot be positive at all value of $r_+$ in the GB statistics (dashed black).} \label{fig:1} 
\end{figure}

The heat capacity $C_R$ is inversely proportional to the slope of the graph $T_R$ versus $r_+$, as shown in Fig.~\ref{fig:1} (left). Namely, we can write
\begin{eqnarray}
C_R = \frac{1}{2}\frac{(1+\Phi^2)}{(1-\Phi^2)^2}\left( \frac{1}{\partial T_R / \partial r_+} \right).
\end{eqnarray}
Therefore, the critical radius $r_c$ is just a turning point at which the slope of $T_R(r_+)$ changes the sign, \textit{i.e.} $T'_R(r_+) = 0$. With the presence of this turning point, the RN-flat black holes with $\lambda = 0.3$ (solid blue), 0.6 (solid green) and 1.0 (solid red) can be stable with positive heat capacity when $r_+>r_c$ and negative heat capacity at $r_+<r_c$, while there is no turning point in the case of GB statistics (dashed black). The disscusion of these results is actually equivalent to the plots in Fig.~\ref{fig:1} (right).

Interestingly, the turning point in Fig.~\ref{fig:1} (left) also shows that the lower bound of a Hawking temperature becomes larger with higher level of non-extensivity. This can be seen through solving \eqref{Tphi} for $r_+$, such that we have
\begin{eqnarray}
r_{L,S} = \frac{2T_R}{\lambda (1-\Phi^2)}\left( 1 \pm \sqrt{1-\frac{\lambda (1-\Phi^2)^2}{4\pi T_R^2}} \right), \label{horizon}
\end{eqnarray}
where $r_L$ and $r_S$ denote the event horizon of the large and small black hole configurations at fixed temperature, respectively. The solution is real when the discriminant in the above equation is greater than zero. This implies that there is the minimum temperature at $r=r_c$, which is 
\begin{eqnarray}
T_{R,min} = \frac{(1-\Phi^2)}{2}\sqrt{\frac{\lambda}{\pi}}.
\end{eqnarray}
Note that for uncharged case, the minimum temperature reduces to the form $T_{R,min} = \frac{1}{2}\sqrt{\frac{\lambda}{\pi}}$ as found in \cite{Czinner1}.

As shown in the work of Gross et al. \cite{Gross}, the 3+1 dimensional hot flat space can be in the state of instability due to the nucleation of black holes without lower bound of temperature. In other words, in the GB statistics, a black hole can be formed at an arbitrarily low temperature of thermal radiation. Considering through R\'enyi statistics, however, gives a different conclusion. A black hole is not allowed to be created when $T_R<T_{R,min}$, as a result from that an event horizon cannot exist due to the presence of an imaginary number in \eqref{horizon}, and the fact that a naked singularity in a 4-dimensional asymptotically flat spacetime is forbidden by the cosmic censorship hypothesis \cite{Penrose}. Accordingly, we can argue that the hot thermal radiation in asymptotically flat spacetime can be collapsed to a black hole only when $T_R>T_{R,min}$. This shows that a Hawking-Page phase transition of black holes in Minkowski background is in a similar way as what occurs in the black holes in AdS background via GB statistics. Therefore, our result is in contrast with that derived from the consideration of Gross et al.~\cite{Gross}.

\begin{figure}
 	\centering
 	 \begin{subfigure}[b]{0.36\textwidth}
 	  \includegraphics[width=\textwidth]{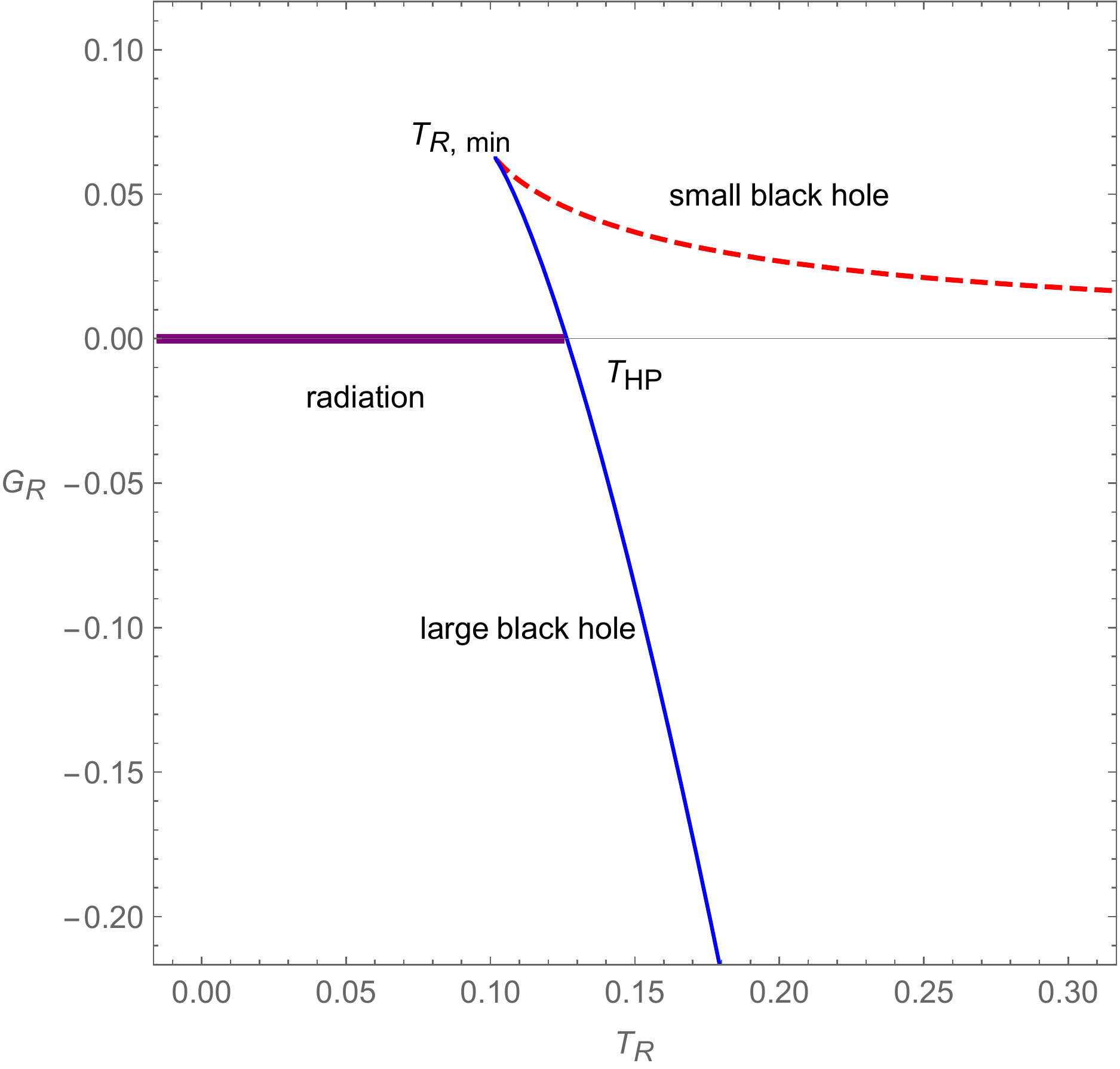}
 	   \end{subfigure}
 	   \begin{subfigure}[b]{0.4\textwidth}
 	    \includegraphics[width=\textwidth]{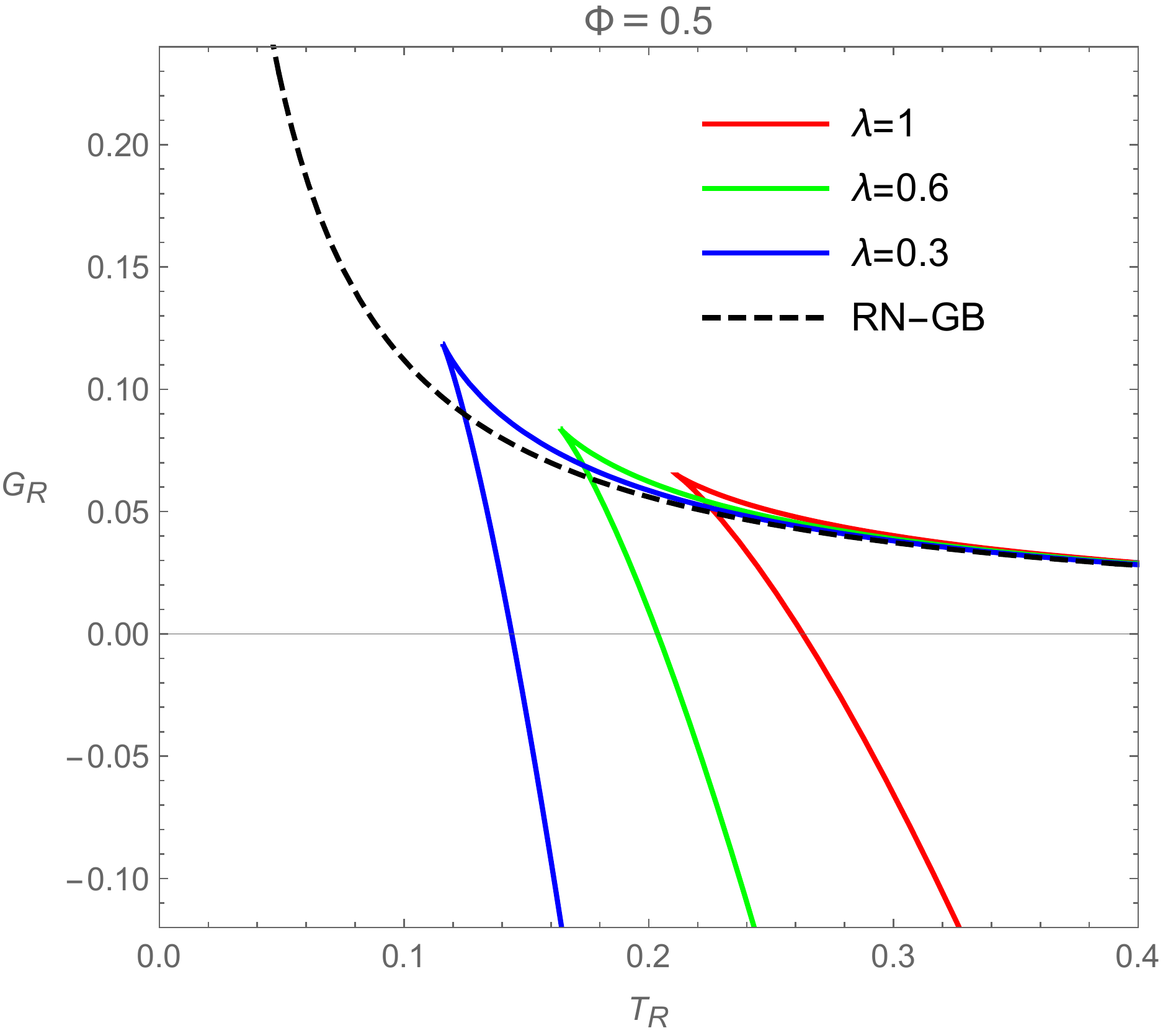}
 	     \end{subfigure}
 	      \caption{(left) The Gibbs free energy vs temperature for fixed $\Phi = 0.5$ in the R\'enyi entropy. (right) The Gibbs free energy vs temperature by varying R\'enyi parameters as $\lambda = 0.3, 0.6,$ and $1$ corresponding to blue, green, and red curves, respectively. The plot from the GB statistics is shown in this graph with black dashed curve.}\label{fig:2}
\end{figure} 

\begin{figure}[h!]
 	\begin{center}
 	\includegraphics[scale=0.5]{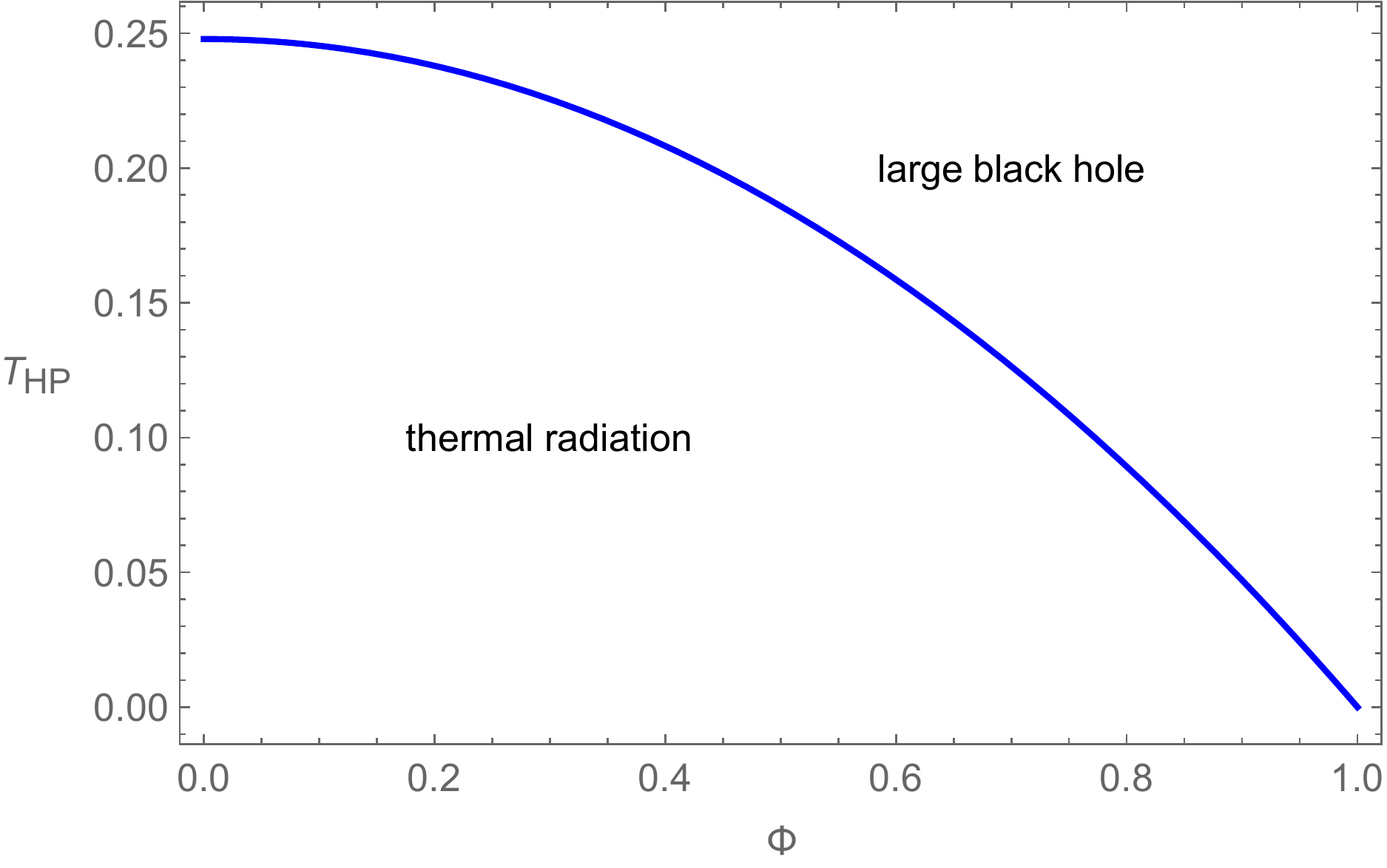}
 	\end{center}
 	 \caption{The Hawking-Page phase transition line in the $T-\Phi$ phase diagram. The line dividing the space into two regions, the thermal radiation phase and large black hole phase.}\label{fig:3}
\end{figure}

The thermodynamic stability can be investigated through considering the free energy as a function of the temperature. In the grand canonical ensemble, we use the Gibbs free energy as thermodynamical potential.  With the R\'enyi statistics, the Gibbs free energy function should be modified by R\'enyi entropy and its corresponding temperature as
\begin{eqnarray}
G_R &=& M - T_{R}S_R - \Phi Q, \nonumber \\ 
       &=& \frac{r_+(1-\Phi^2)}{4}\left[ 2 - \left( 1 + \frac{1}{\lambda \pi r_+^2} \right)\ln (1+\lambda\pi r_+^2) \right]. \label{free}
\end{eqnarray}
The plots of $G_R$ versus $T_R$ are shown in Fig.~\ref{fig:2} (left). While the GB statistics ($\lambda=0$) gives us that the charged black hole is unstable as its free energy is positive at any non-zero temperature (dashed, Fig.~\ref{fig:2} (right)), the R\'enyi parameter ($\lambda >0$), allows that the black hole can be stable above a certain temperature $T_{HP}$, where the $G_R$ has negative value as indicated in Fig.~\ref{fig:2} (left). From \eqref{free}, we can find this temperature at which the first order Hawking-Page phase transition occurs by setting $G_R = 0$. Numerically, we can find that 
\begin{eqnarray}
T_{HP} \approx 0.64(1-\Phi^2)\sqrt{\frac{\lambda}{\pi}},
\end{eqnarray}
which is about 1.28 times $T_{R,min}$. Fig.~\ref{fig:3} shows the Hawking-Page temperature as a function of the electrostatics potential. This phase transition behavior from our results are very similar to the RN-AdS via standard GB statistics \cite{Chamblin1,Chamblin2,Chatchai}. Obviously, we can see that for $\Phi=1$ and $\lambda=0$ cases the Hawking-Page temperature vanishes, therefore the first order phase transition cannot exist either in the extremal black hole or the consideration of black holes via the GB statistics.

Let us focus on the presence of the cusp of $G_R$ at $T_{R,min}$ which is less than $T_{HP}$. The cusp corresponds to the point of small/large black hole phase transition. This is of the second order type due to the discontinuity of the second derivative of the Gibbs free energy with respect to the temperature at the cusp, as seen clearly through the divergence of heat capacity, $C_R = -T_R\left( \frac{\partial^2G_R}{\partial T_R^2} \right)_\Phi$, at the cusp as shown in Fig.~\ref{fig:2}.
\begin{eqnarray}
S_{Rc} = \frac{1}{\lambda}\ln 2 \ , \ \ \ G_{Rc} = \frac{(1-\Phi^2)}{2\sqrt{\lambda \pi}}(1-\ln 2).
\end{eqnarray}
Notice that $S_{Rc}$ is a universal constant at a fixed $\lambda$.  It is obviously independent of black hole's mass $M$ and charge $Q$. At the cusp, the Gibbs free energy is at the highest value $G_{Rc}$, as shown in the equation above. 

In the AdS/CFT corespondence, the electric potential $\Phi$ in the bulk has the holographic dual to the chemical potential $\mu$ of the gauge theory at the boundary. The Hawking-Page phase transition in the bulk of AdS space is dual to the phase transition between a cold confined phase and a hot deconfined phase at finite chemical potential. Intriguingly, the phase diagram from our results, as shown in Fig.~\ref{fig:3}, looks similar to the $T-\mu$ phase diagram of confining/deconfining phase transition at finite chemical potential. This provides one more supporting evidence that physics of asymptotically flat black hole using R\'enyi approach somehow corresponds to that of asymptotically AdS black hole using GB statistics, as suggested in previous works \cite{Czinner1,Czinner2}.

\subsection{Canonical Ensemble}

\begin{figure}
 	\centering
 	 \begin{subfigure}[b]{0.41\textwidth}
 	  \includegraphics[width=\textwidth]{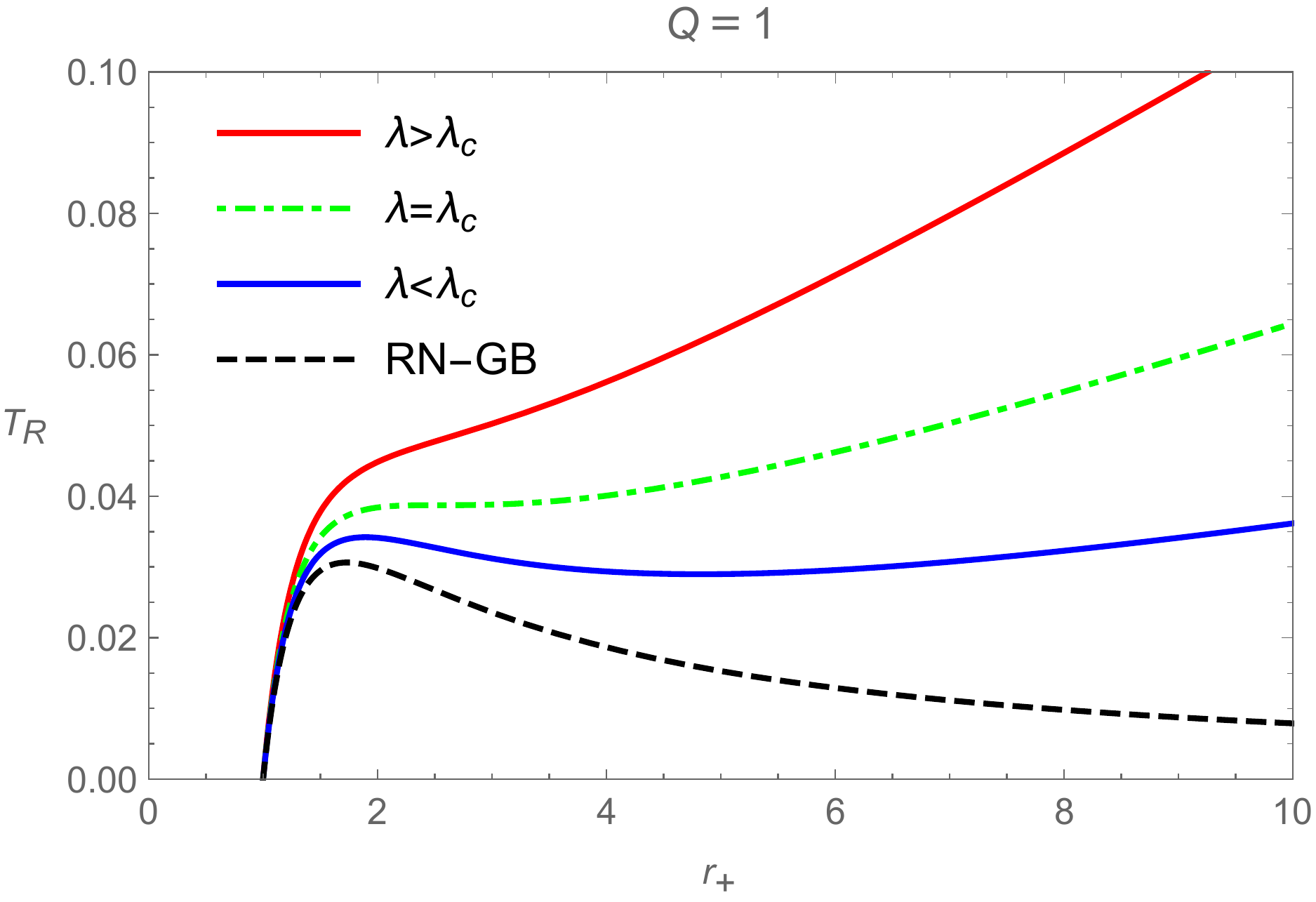}
 	   \end{subfigure}
 	   \begin{subfigure}[b]{0.58\textwidth}
 	    \includegraphics[width=\textwidth]{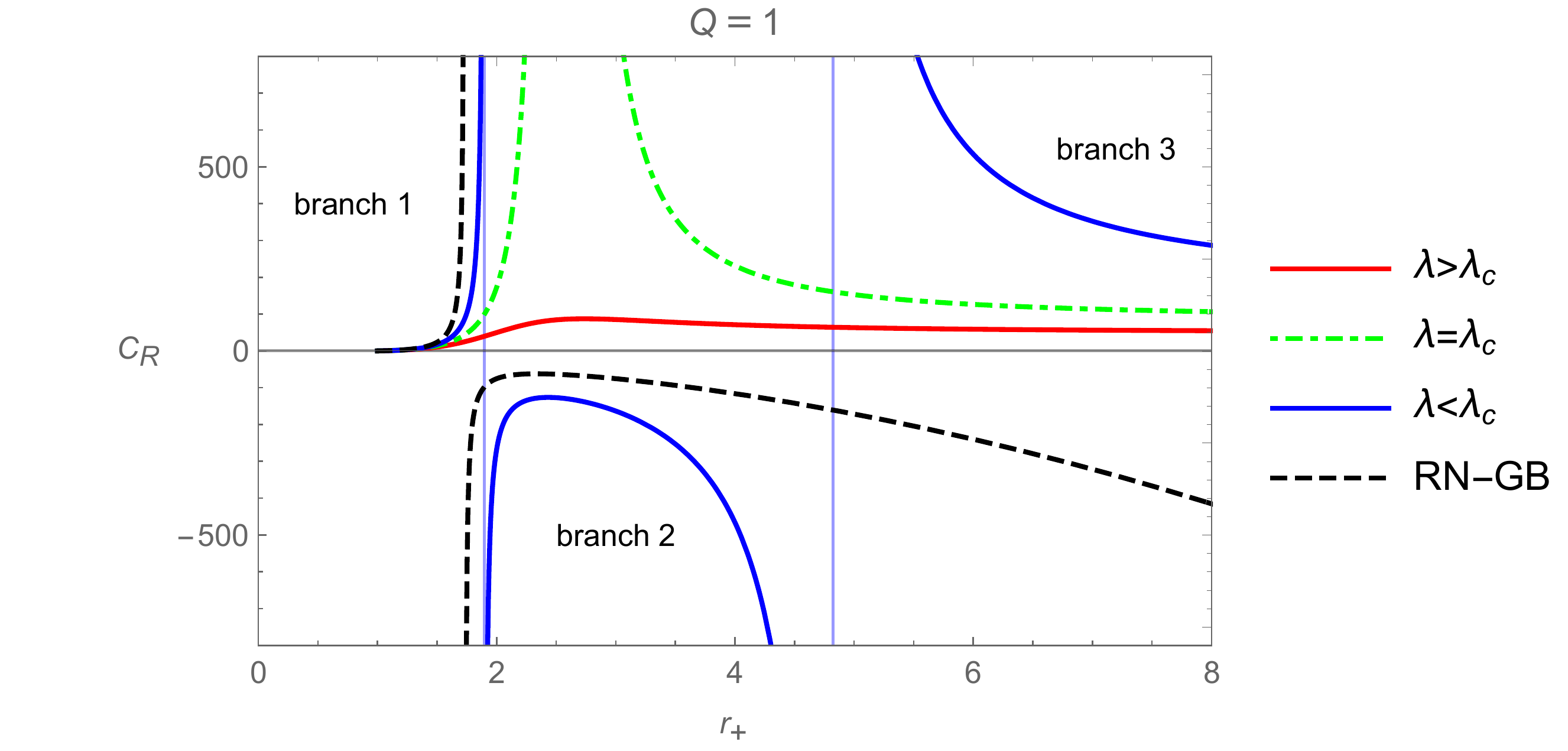}
 	     \end{subfigure}
 	      \caption{(left) The plots of the R\'enyi temperature of charge black hole $T_R$ versus the event horizon radius $r_+$ with fixed charge $Q=1$ at different value of parameter $\lambda$. (right) The heat capacity $C_R$ versus $r_+$ with charge $Q=1$ at different value of $\lambda$. }\label{fig:4}
\end{figure}

Here, we investigate the thermal properties and the phase structure of  black holes in canonical ensemble (fixed $Q$) via R\'enyi statistics. Using \eqref{Thq} and \eqref{Tr} the Hawking temperature with fixed charge can be written as
\begin{equation}
T_{R} = \frac{(r_+^{2}-Q^2)(1+ \lambda \pi r_+^{2})}{4\pi r_+^{3}}. \label{bh25}
\end{equation}
The relation between $T_R$ and $r_+$ with fixed $Q=1$ at different $\lambda$ are plotted as shown in Fig.~\ref{fig:4} (left). These isocharge curves of the black hole temperature versus the horizon radius look similar to the case of non-zero charge black holes in AdS background \cite{Chamblin1,Chamblin2,Chatchai}. For different values of $\lambda$ parameter, the isocharge curves behave differently where a critical phase transitioin can occur with the condition
\begin{equation}  
0=\left(\frac{\partial T_{R}}{\partial r_+}\right)_Q=\left(\frac{\partial^2 T_{R}}{\partial r_+^2}\right)_Q. \label{bh26}
\end{equation}
By solving this equation, we obtain the critical Renyi parameter $\lambda_c$, critical horizon $r_c$ and critical temperature $T_c$ at the critical point as
\begin{eqnarray}
\lambda_c &=& \frac{7 - 4\sqrt{3}}{\pi Q^2}, \\
r_c &=& (3+2\sqrt{3})^{1/2}Q, \\
T_c &=&  \frac{2}{(3+2\sqrt{3})^{3/2}\pi Q}.
\end{eqnarray}
For small R\'enyi parameter $\lambda<\lambda_c$, there are two local extrema of the isocharge curve in Fig.~\ref{fig:4} (left) at
\begin{eqnarray}
r^2_{1,2} = \frac{1}{2\pi \lambda}\left[ (1-\lambda \pi Q^2) \mp \sqrt{\lambda^2\pi^2Q^4 -14\lambda \pi Q^2 + 1} \right], \label{r12}
\end{eqnarray}
where the horizon radius $r_1$ and $r_2$ correspond to a local maximum and local minimum of the isocharge curve, respectively. The discriminant in \eqref{r12} is zero as $\lambda=\lambda_c$, hence two extremal radii $r_1$ and $r_2$ are degenerate into $r_c$.

The corresponding heat capacity is
\begin{eqnarray}
C_{R} &=& T_{R}\left( \frac{\partial S_{R}}{\partial T_{R}} \right)_{Q} = -\frac{2\pi r_+^2(r_+^2-Q^2)}{r_+^2-3Q^2+\lambda \pi r_+^2(r_+^2+Q^2)}. \label{C}
\end{eqnarray}
In contrast with GB statistics, the R\'enyi approach allows three branches of charged black hole configurations in canonical ensemble, one negative and two positive heat capacity as shown in Fig.~\ref{fig:4} (right). We denote them as branch 1, branch 2 and branch 3, respectively. The additional branch 3 now appears and shows an interesting thermal phase of black hole. From \eqref{C}, the heat capacity grows without upper bound when the horizon radius $r_+$ equals $r_1$ and $r_2$. The coresponding temperature at radius $r_1$ and $r_2$ are $T_1=0.0299$ and $T_2=0.0296$, respectively, for the chosen parameters in the Fig.~\ref{fig:4}.  
\begin{figure}
 	\centering
 	 \begin{subfigure}[b]{0.44\textwidth}
 	  \includegraphics[width=\textwidth]{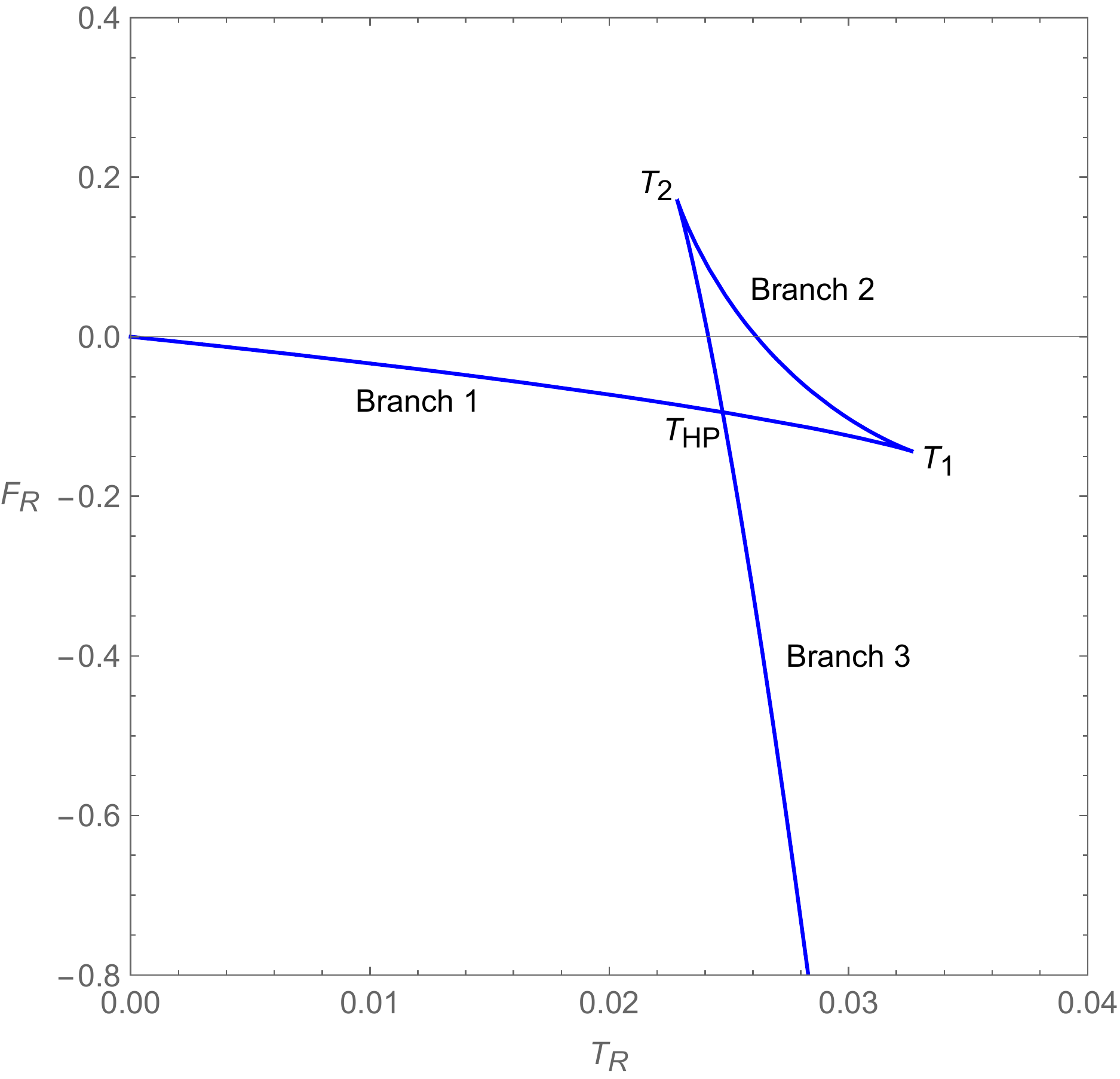}
 	   \end{subfigure}
 	   \begin{subfigure}[b]{0.43\textwidth}
 	    \includegraphics[width=\textwidth]{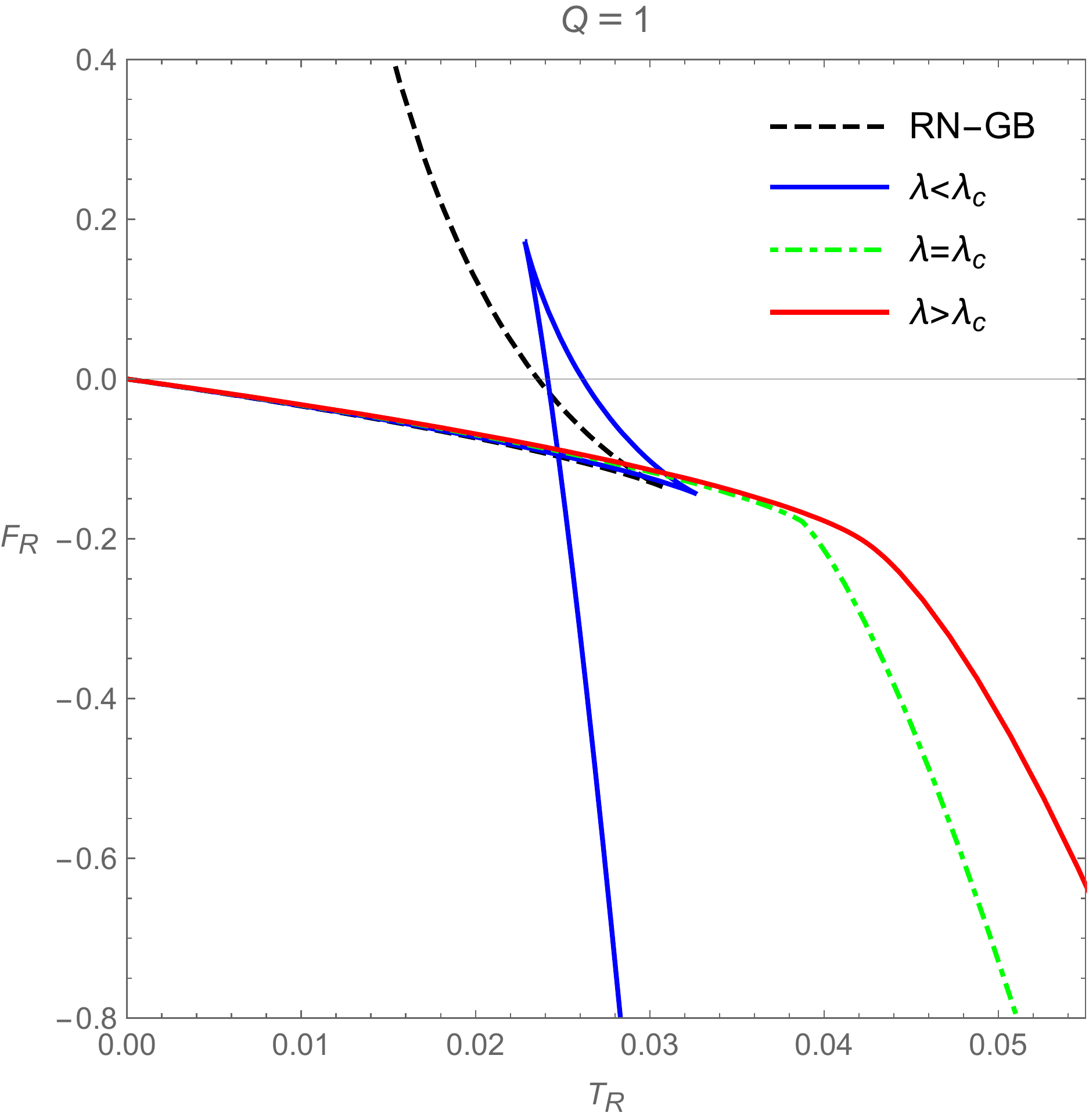}
 	     \end{subfigure}
 	      \caption{Plot of the Helmholtz free energy of a charged black hole in a fixed charged case at $Q = 1$ against the Hawking temperature.}\label{fig:5}
\end{figure}
In the canonical ensemble, the thermodynamic potential is the Helmholtz free energy which can be generalized to satisfy the R\'enyi statistics as
\begin{eqnarray}
F_R &=& E - T_{R}S_R, \nonumber \\
&=&\frac{Q-r_+}{4 \pi r_+^3}\left[ 2\pi r_+^2(Q-r_+) + \frac{1}{\lambda}(Q+r_+)(1+\lambda \pi r_+^2)\ln (1+\lambda \pi r_+^2) \right], 
\end{eqnarray}
where $E = M - M_e$ is the energy of the system relative to that of the extremal black hole. Fig.~\ref{fig:5} (left) shows the relation of the free energy versus the R\'enyi temperature of RN-flat black holes at a small $\lambda$ parameter. Interestingly, the swallowtail behavior occurs when $\lambda < \lambda_c$, which is very similar to a Van der Waals type of liquid/gas phase transition. In Fig.~\ref{fig:5} (left), there is only the branch 1 that exists at low temperature. At a certain temperature $T_2$, the branch 2 and 3 emerge simultanously with larger free energy than the branch 1. However, when the temperature increases to $T_1$, it can be seen that the branch 1 and 2 combine and disappear. Just a little bit below $T_1$, the free energy of the branch 1 and 3 are equal at temperature $T_{HP}$ and the latter has more negative free energy than the first at $T>T_{HP}$. This result implies the Hawking-Page phase transition from branch 1 to branch 3 at this point. Furthermore, for $\lambda > \lambda_c$ there is only one large black hole configuration with positive heat capacity, therefore it is thermodynamically stable. The graphs of the free energy versus R\'enyi temperature at different values of $\lambda$ is shown in Fig.~\ref{fig:5} (right). 

\section{The Emergence of Van der Waals Phase Transition}

In the extended phase space approach with the GB statistics, a complete analogy between VdW liquid and RN-AdS black holes in canonical ensemble was established by identifying the thermodynamic pressure with cosmological constant, $P=-\frac{\Lambda}{8\pi}$. However, in the asymptotically flat spacetime, the thermodynamic pressure $P=0$ and the black hole phase transition in the $F-T$ plane is not in the VdW type. As discussed in the previous section, the $\lambda$ parameter can give an effect to the thermodynamic description in the same way as the existence of the negative cosmological constant $\Lambda$. {In this section, we will derive a consistent Smarr formula for R\'enyi statistics and show the results that the phase structure of RN-flat black hole through \textsl{R\'enyi extended phase space} can have the behavior of the VdW liquid-gas system in the canonical ensemble.}

\subsection{Generalized Smarr Formula}      

To achieve this, we will extend the usual Smarr formula in term of the R\'enyi entropy and its corresponding temperature instead of Bekenstein-Hawking entropy and Hawking temperature. In four-dimensional RN-flat black holes, the Smarr formula is 
\begin{eqnarray}
M = 2T_HS_{BH} + \Phi Q. \label{Smarr}
\end{eqnarray}
Substituting $T_H=\frac{T_R}{e^{\lambda S_R}}$ and $S_{BH}=\frac{e^{\lambda S_R} - 1}{\lambda}$ derived from \eqref{bh17} and \eqref{Tr} into \eqref{Smarr} and expanding the first term in the power series of $\lambda$ parameter, we obtain  
\begin{eqnarray}
M &=& 2T_RS_R - \lambda T_RS_R^2+\Phi Q + O(\lambda^2), \nonumber \\
&=& 2T_RS_R - \frac{\lambda (1-\Phi^2)}{4}\pi r_+^3 + \Phi Q + O(\lambda^2). \label{Low}
\end{eqnarray}
To get the second term of the last line, we have substituted $T_R$ from \eqref{Tphi} and expanded $S_R = \pi r_+^2 - \frac{1}{2}\lambda \pi^2 r_+^4 + O(\lambda^2)$. Then, from the second term of \eqref{Low}, we can identify 
\begin{eqnarray}
P = \frac{3\lambda (1-\Phi^2)}{16}, \ \ \ V=\frac{4}{3}\pi r_+^3. \label{P}
\end{eqnarray} 
Therefore, the lowest order approximation {of $\lambda$} in \eqref{Low} can be written in the form
\begin{eqnarray}
M = 2T_RS_R - PV + \Phi Q. 
\end{eqnarray}
We call this relation here as a modified Smarr formula via R\'enyi statistics. Obviously, we have suggested that the new term in the Smarr formula is the product of thermodynamic pressure $P$ and the thermodynamic volume $V$. The thermodynamic volume is a conjugate quantity to the pressure, which can be obtained through the standard relation $V=\left(\frac{\partial M}{\partial P}\right)_{\Phi ,S_R}$ at leading order of $\lambda$. Hence it is not a geometric spherical volume with horizon radius $r_+$. Our result shows that the mass of black hole $M$ should be interpreted as the enthalpy $H$ rather than the internal energy of the system in a similar way to the results from the extended phase space approach \cite{Kastor}.

\subsection{The Equation of State}  

Typically in a thermodynamic system, the matter changes its temperature when their microscopic components emit or absorb photons. In the same way, a black hole can change its temperature through gaining or losing their masses. By this analogy, we may assume that the black hole could have their own microstates carrying the degrees of freedom as well. To describe the black hole microscopic structure, each individual microstate of black hole has been proposed to contain $\gamma$ Plank area pixels of event horizon surface \cite{Wei}. Thus, the total number of degrees of freedom is
\begin{eqnarray}
N = \frac{A}{\gamma l_p^2}, \label{N}
\end{eqnarray}
where $l_p=\sqrt{\frac{\hbar G}{c^3}}$ is Plank length and $l_p^2$ is the area of one Plank area pixel. We can say that each individual constituent of a black hole has the \textsl{specific volume}. Using \eqref{P} and \eqref{N}, the black hole has  the specific volume of the form 
\begin{eqnarray}
v = \frac{V}{N} = \frac{\gamma l^2_p}{3}r_+. \label{volume}
\end{eqnarray}  
Note that the specific volume $v$ scales linearly in the horizon radius $r_+$ of the black holes.  In the remaining of this section, we will explore about the equation of state and thermal phase diagram from this approach in both grand canonical and canonical ensembles.
 
\subsubsection{Grand Canonical Ensemble}

\begin{figure}
 	\centering
 	 \begin{subfigure}[b]{0.4\textwidth}
 	  \includegraphics[width=\textwidth]{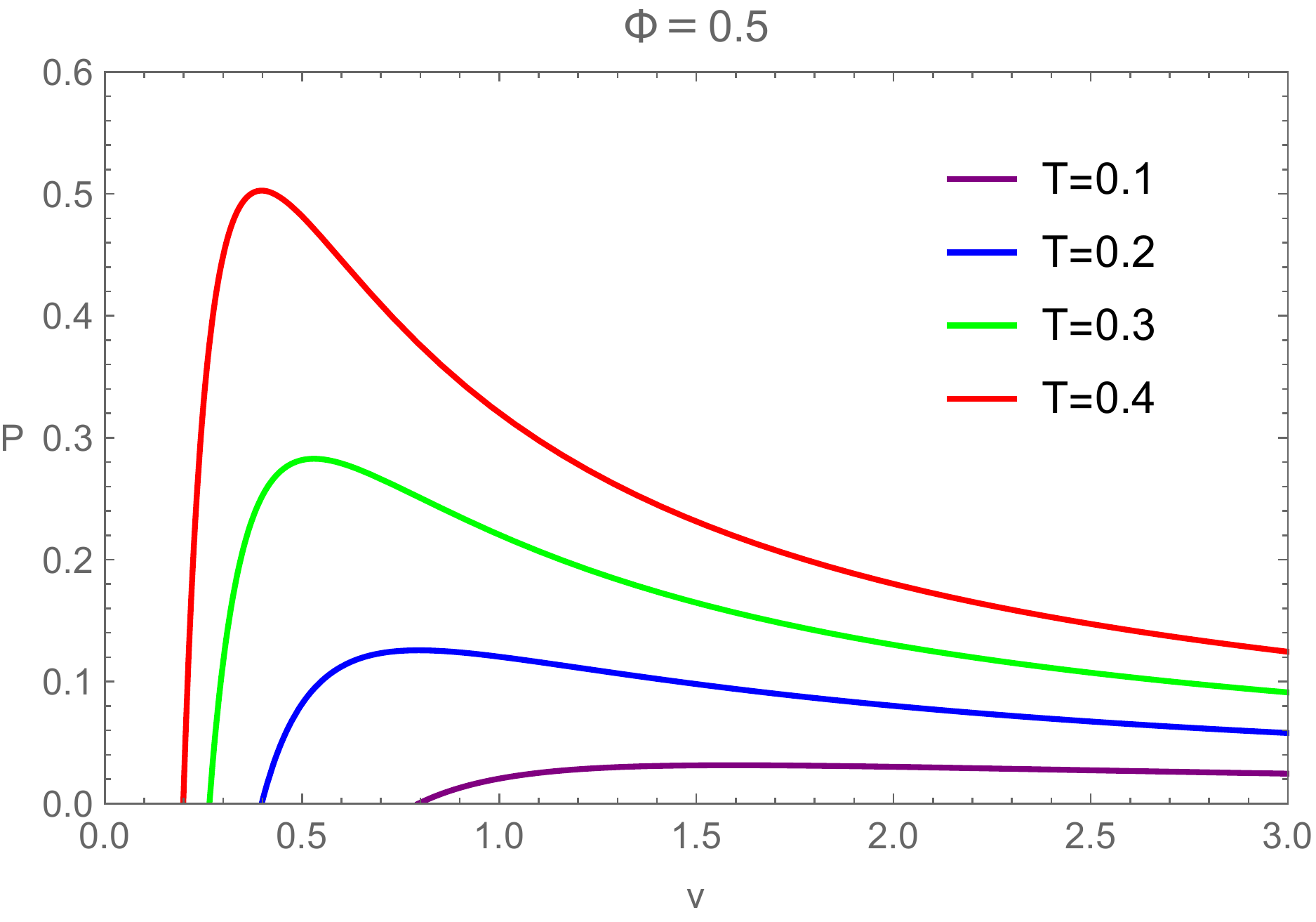}
 	   \end{subfigure}
 	   \begin{subfigure}[b]{0.4\textwidth}
 	    \includegraphics[width=\textwidth]{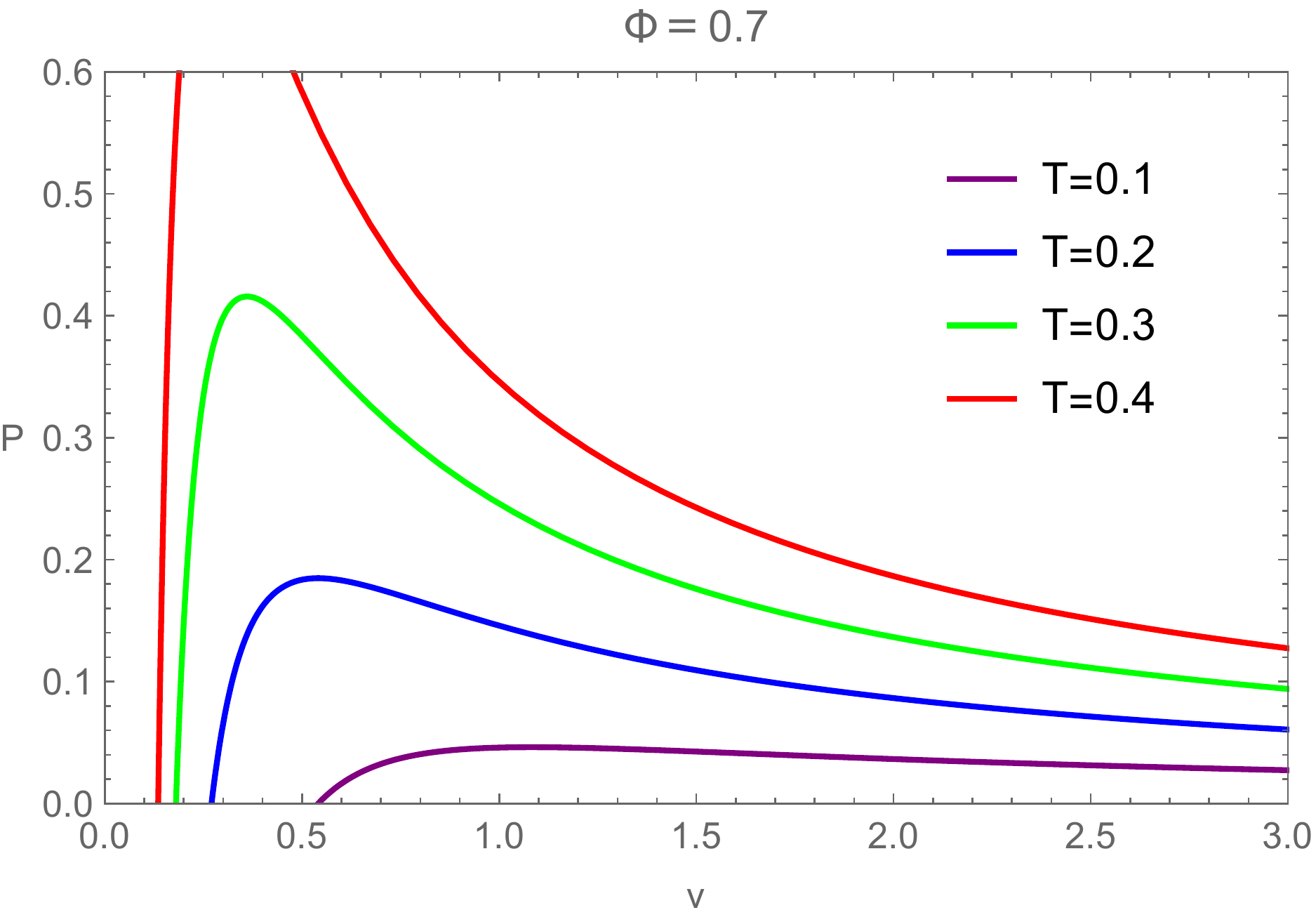}
 	     \end{subfigure}
 	      \caption{$P-v$ diagram of charged black holes for fixed potential $\Phi = 0.5$ (left) and $\Phi = 0.7$ (right). In both graphs, the temperature of isotherms decreases from top to bottom. For a given temperature, there are two branches of black hole, one is the small black hole ($P$ increases as $v$ increases) and the other is the large black hole ($P$ decreases as $v$ increases).}\label{fig:6}
\end{figure}

With the macroscopic perspective, the equation of state of RN-flat black hole in the fixed $\Phi$ ensemble can be obtained from the formula of $P, v, T_R$. Summarizing from \eqref{Tphi}, \eqref{P} and \eqref{volume}, we now have
\begin{eqnarray}
T_{R}  =  \frac{(1-\Phi^2)(1+\lambda \pi r_+^{2})}{4\pi r_+}, \ \ \ P = \frac{3\lambda (1-\Phi^2)}{16}, \ \ \ v=\frac{\gamma l^2_p}{3}r_+. \label{pvphi} 
\end{eqnarray}
Combining these equation with the elimination of $\lambda$, we will arrive at the equation of state 
\begin{eqnarray}
P = \frac{T_R}{v} - \left(\frac{1-\Phi^2}{3\pi}\right)\frac{1}{v^2}, \label{eosphi}
\end{eqnarray}
where we have used $l_p = 1$ and $\gamma = 4$ in \eqref{eosphi}. Remark that the equation of state is written in the form of the specific volume $v$, which is proportional to $r_+$ rather than thermodynamic volume $V$. In the grand canonical ensemble, the relations between $P$ and $v$ in isothermal process with different values of $T_R$ are plotted at $\Phi=0.5, 0.7$ in Fig.~\ref{fig:6}.  By solving the condition $\left(\frac{\partial P}{\partial v}\right)_{T_R,\Phi}=0$, the pressure has a maximum value $P_{max}$ at $v_c$ as 
\begin{eqnarray}
P_{max}=\frac{3\pi T^2_R}{4(1-\Phi^2)}, \ \ \ v_c=\frac{2(1-\Phi^2)}{3\pi T_R}. \label{Pmax}
\end{eqnarray}
For a given temperature, there are two black hole configurations below a maximum pressure $P_{max}$ corresponding with small and large black holes when $v<v_c$ and $v>v_c$, respectively. {The small black hole phase is unstable because $P$ increases as 
$v$ increases, therefore the compression coefficient is negative. In the large black hole phase, the compression coefficient is positive because $P$ decreases as $v$ increases, hence it is thermodynamically stable.} We can find an \textsl{ideal gas} behavior for large black holes at high temperature. However, as seen from \eqref{Pmax}, the second derivative $\left(\frac{\partial^2P}{\partial v^2}\right)_{T_R,\Phi}$ cannot vanish at $v_c$. This means that there is no critical behavior for RN-flat black hole in the grand canonical ensemble.

\subsubsection{Canonical Ensemble}

For the canonical ensemble, it has been known that the transition of small/large RN-AdS black holes in extended phase space analogous to the liquid/gas phase transition of the VdW type. Therefore, it is interesting to see whether the $P-v$ criticality appears or not for R\'enyi extended phase space approach in this ensemble. As in the previous section, we identify the maccroscopic quantities as 
\begin{eqnarray}
T_{R} = \frac{(r_+^{2}-Q^2)(1+ \lambda \pi r_+^{2})}{4\pi r_+^{3}}, \ \ \ P=\frac{3\lambda}{16}\left( 1- \frac{Q^2}{r_+^2}\right), \ \ \ v=\frac{4}{3}r_+, \label{pvq} 
\end{eqnarray}
where we have used $\Phi=\frac{Q}{r_+}$, $l_p=1$ and $\gamma=4$. The equations of state in term of $P$, $v$ and $T_R$ can be written in the form
\begin{figure}
 	\centering
 	 \begin{subfigure}[b]{0.4\textwidth}
 	  \includegraphics[width=\textwidth]{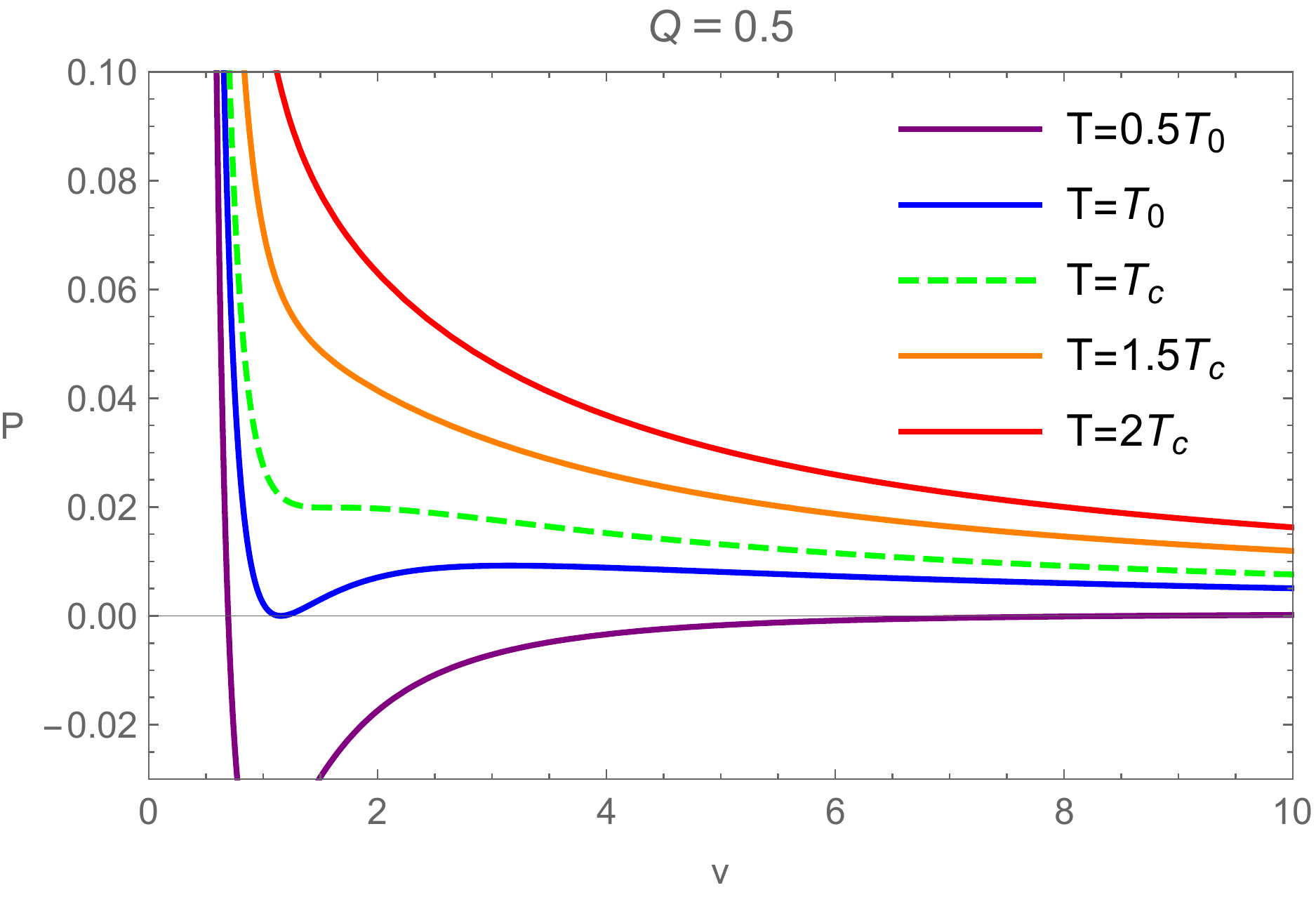}
 	   \end{subfigure}
 	   \begin{subfigure}[b]{0.4\textwidth}
 	    \includegraphics[width=\textwidth]{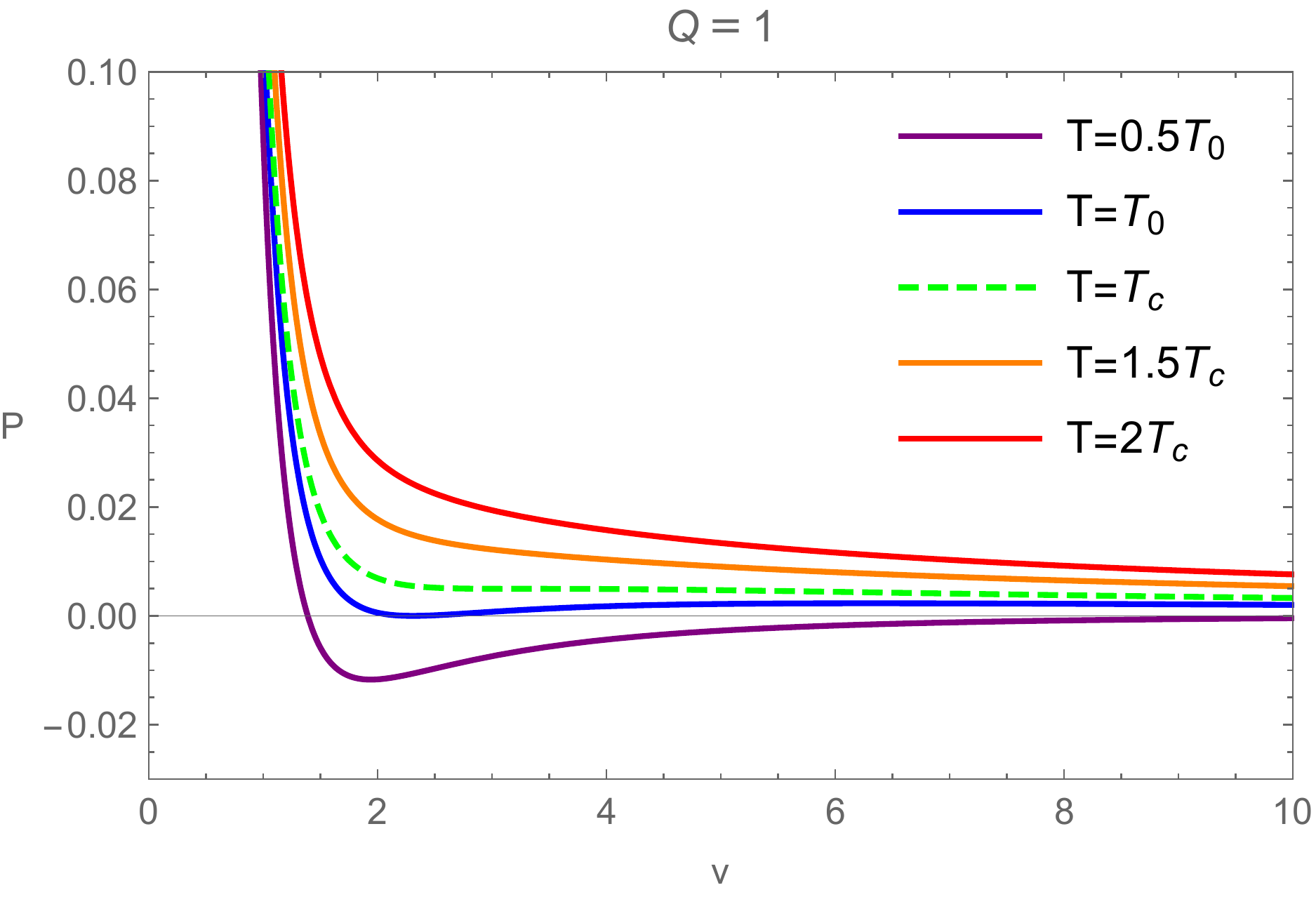}
 	     \end{subfigure}
 	      \caption{$P-v$ diagram of charged black holes for fixed charge $Q=0.5$ (left) and $Q=1$ (right). In both graphs, the temperature of isotherms decreases from top to bottom. The dashed green line is denoted as the critical isotherm $T=T_c$. The lower solid lines correspond to small temperature $T<T_c$, which have three branches of black holes, one unstable ($P$ increases as $v$ increases) and two stable ($P$ decreases as $v$ increases) configurations. The upper solid lines are in high temperature limit $T>T_c$, analogous to the ideal gas phase of VdW system.}\label{fig:7}
\end{figure}
\begin{eqnarray}
P = \frac{T_R}{v} - \frac{1}{3\pi v^2} + \frac{16Q^2}{27\pi v^4}.
\end{eqnarray}
At a fixed value of $Q$, this formula indicates that the $P-v$ diagram shows a critical behavior like the VdW liquid-gas system, as shown in Fig.~\ref{fig:7}.  The phase structure can be critically changed when the temperature crosses the critical point, which is the point of inflection with these two conditions 
\begin{eqnarray}
\left(\frac{\partial{P}}{\partial v}\right)_{T_R,Q}=0, \ \ \ \left(\frac{\partial ^2{P}}{\partial v^2}\right)_{T_R,Q}=0.
\end{eqnarray}
Solving these conditions lead to the critical thermodynamic parameters of the charged black holes as follows
\begin{eqnarray}
P_c&=&\frac{1}{64\pi Q^2}, \\
v_c&=& 4\sqrt{\frac{2}{3}}Q , \\
T_c&=&\frac{1}{3\sqrt{6}\pi Q}.
\end{eqnarray}
These formula of critical values lead us to the critical compressibility factor
\begin{eqnarray}
Z_c \equiv \frac{P_cv_c}{T_c}=\frac{3}{8},
\end{eqnarray}
which is a universal constant in the sence that it is independent of black hole's mass and charge. Interestingly, this universal constant from our results has the same value as in the VdW fluid and RN-AdS black hole in the extended phase space \cite{Mann1}. {Remark that $Z_c=\frac{3}{8}$ exactly when the factor $\gamma=4$, this indicates that we have four Plank area pixels to carry one degree of freedom in the R\'enyi extended phase space model. Therefore the value of the critical compressibility factor depends on the number of Plank area pixels which represents one degree of freedom for black holes. For the RN-AdS black hole, it is necessary to choose $\gamma=6$ to obtain the result $Z_c=\frac{3}{8}$.} 

The $P-v$ diagram in Fig.~\ref{fig:7} is the same as in the VdW liquid-gas system. In the left and right graphs of this figure corespond to the case of fixed charge $Q=0.5, 1$, respectively. The critical isothermal curve is depicted by the green dashed line. For $T_R<T_c$, there is a small black hole phase at small value of $v$, corresponding to liquid phase, while there is a large black hole phase at large value of $v$, corresponding to gas phase. These two stable configurations have a positive compression coefficient ($P$ decreases as $v$ increases).  We also have the unstable region with negative compression coefficient ($P$ increases as $v$ increases) between these two regions at which the black hole of intermediate size can represent the mixture of liquid and gas phases. Moreover, there is a minimum temperature $T_0$, given by
\begin{eqnarray}
T_0 = \frac{1}{6\sqrt{3}\pi Q},
\end{eqnarray}
associated with $P=0$ at $v_0=\frac{4Q}{\sqrt{3}}$, as seen in the blue solid line in Fig.~\ref{fig:7}. For $T<T_0$, the black hole pressure becomes negative at some specific volume $v$. In a high temperature case $T_R>T_c$, the black holes behave like an ideal gas and no phase transition occur.

\section{Conclusion}

In this paper, we investigate the thermodynamics of charged black holes in asymptotically flat spacetime or RN-flat via an alternative R\'enyi entropy both in the grand canonical and canonical ensembles. Applying the R\'enyi statistics, our results indicate that it is possible, with non-zero $\lambda$, to have the small and large black hole branches in the grand canonical ensemble, while this cannot occur in the GB statistics approach. Moreover, we have shown that the Hawking-Page phase transition between thermal radiation and large black hole in RN-flat can happen and crucially depend on $\lambda$ parameter.

For the canonical ensemble, in contrast with the grand canonical case, there is a critical behavior, such that, when the R\'enyi parameter $\lambda$ is less than the critical value $\lambda < \lambda_c$, a small/large black hole first order phase transition can occur.  Above the critical value of $\lambda$, this phase transition disappears and the large black hole phase is only the remaining phase that is possible. This thermal behavior is in the same way as that of the charged black hole in AdS space with the standard GB statistics.

A consistent Smarr formula for R\'enyi statistics has been derived. We show that the $\lambda$ parameter will contribute to the thermodynamic pressure for black holes. Then the quantity conjugate to the pressure may be quantified as a thermodynamic volume of black holes. This results have shown that the black hole mass represents the enthalpy rather than the internal energy. After  introducing a specific volume via number of degrees of freedom and thermodynamic volume of black holes, we have tried to write the equation of state and study the thermal behavior of charged black holes in flat space with non-zero $\lambda$ parameter. In the $P-v$ diagram, we observe that its thermal phase structure in canonical ensemble is analogous to the VdW phase transition. Especially, taking $\gamma=4$, we will obtain the critical compressibility factor $Z_c=\frac{3}{8}$ as in the VdW equation of state.

\section*{Acknowledgement}
We are grateful to Pitayuth Wongjun, Supakchai Ponglertsakul, Sirachak Panpanich and Krai Cheamsawat for discussions. This work has been supported by the Petchra Pra Jom Klao Ph.D. Research Scholarship from King Mongkut's University of Technology Thonburi.

\end{document}